\documentstyle[epsfig, rotating]{jaa}

\DeclareMathAlphabet{\mathsc}{OT1}{cmr}{m}{sc}
\def\testbx{bx}%
\DeclareRobustCommand{\ion}[2]{%
\relax\ifmmode
\ifx\testbx\f@series
{\mathbf{#1\,\mathsc{#2}}}\else
{\mathrm{#1\,\mathsc{#2}}}\fi
\else\textup{#1\,{\mdseries\textsc{#2}}}%
\fi}

\def\HII{\ion{H}{ii}~} 
\def\HeI{\ion{He}{i}~}
\def\HeII{\ion{He}{ii}~}
\def\CIV{\ion{C}{iv}~}  
\def\NII{\ion{N}{ii}}
\def\SII{\ion{S}{ii}}
\def\Ha{H$\alpha$~}

\def\deg{\hbox{$^\circ$}~}

\begin{document}
\title[Mrk~996 using DFOT]
{Devasthal Fast Optical Telescope observations of Wolf-Rayet dwarf galaxy Mrk~996}
\author[Jaiswal \& Omar]
{S. Jaiswal$^{1}$\thanks{e-mail: sumit@aries.res.in}
\& A. Omar$^{1}$\thanks{e-mail: aomar@aries.res.in}\\
(1) Aryabhatta Research Institute of Observational Sciences (ARIES), 
Manora Peak, Nainital, 263002, India
}

\pubyear{xxxx}
\volume{xx}
\date{Received xxx; accepted xxx}
\maketitle
\label{firstpage}
\begin{abstract}

The Devasthal Fast Optical Telescope (DFOT) is a 1.3 meter aperture optical 
telescope, recently installed at Devasthal, Nainital. We present here the first 
results using an \Ha filter with this telescope on a Wolf-Rayet dwarf galaxy 
Mrk~996. The instrumental response and the \Ha sensitivity obtained with the 
telescope are $(3.3 \pm 0.3)\times 10^{-15}~(erg~s^{-1}~cm^{-2})/(counts~s^{-1})$ 
and $7.5\times10^{-17}~erg~s^{-1}~cm^{-2}~arcsec^{-2}$ respectively. The \Ha flux 
and the equivalent width for Mrk~996 are estimated as 
$(132 \pm 37)\times 10^{-14}~erg~s^{-1}~cm^{-2}$ and $\sim$96~$\mathring{A}$ 
respectively. The star formation rate is estimated as  
$0.4\pm0.1~M_\odot$yr$^{-1}$. Mrk~996 deviates from the radio-FIR correlation 
known for normal star forming galaxies with a deficiency in its radio continuum. 
The ionized gas as 
traced by \Ha emission is found in a disk shape which is misaligned with 
respect to the old stellar disk. This misalignment is indicative of  a recent 
tidal interaction in the galaxy. We believe that galaxy-galaxy tidal 
interaction is the main cause of the WR phase in Mrk~996.

\end{abstract} 


\section{Introduction}

Wolf-Rayet (WR) galaxies are a special type of star-burst galaxies where 
an episode of massive star formation is only a few Myr old (Schaerer et al.\ 
1999). These galaxies are a subset of \HII galaxies, whose integrated 
spectra show broad emission lines (mainly \HeII 4686 and \CIV 5808; also N and 
O) attributed to a presence of WR stars (Conti 1991). The 
strengths of these emission lines indicate a substantial population ($10^2$ - 
$10^5$) of WR stars. The most massive O-type stars ($M \ge 25 M_{\odot}$) go 
through the WR phase a few Myr after their birth, spending 
only a short time ($t_{WR} \le 1$ Myr) in this phase until they explode as 
supernovae (Meynet \& Maeder 2005). Therefore, WR galaxies offer an opportunity 
to study phenomena associated with the very early phases as well as the triggering 
mechanisms of massive star-formation in galaxies. WR galaxies are quite rare in 
nearby universe essentially because of the short life-time of WR stars. 
The most extensive catalog to date of WR galaxies, containing 570 galaxies was 
compiled by Brinchmann et al.\ (2008) based on a search through the SDSS (Sloan 
Digital Sky Survey) data release 6. These authors find 
that WR phase is found in almost all the morphological types ranging from 
low-mass, blue compact low-metallicity dwarf galaxies to massive spirals, 
luminous mergers, IRAS galaxies and Seyfert galaxies.

In this paper, \Ha observations of Mrk~996 are presented using the newly 
installed 1.3m Devasthal Fast Optical Telescope (DFOT) near Nainital in India. 
Mrk~996 ($\alpha,~\delta~(J2000) = 01^{h}~27^{m}~35.5^{s},~
-06^{d}~19^{m}~36^{s}; V_{rad-Hel}= 1622 \pm 10$~km~s$^{-1}$) belongs to a 
relatively rare class of nE BCDs (Blue Compact Dwarfs). The nE BCDs 
have a clearly defined nucleus (n) within an elliptical (E) halo. The BCDs are galaxies 
fainter than $M_B= -18.15$ mag ($H_0=70 km~s^{-1}~Mpc^{-1}$), with size less than 
1 kpc, and having strong emission lines superposed on a blue continuum (Thuan \& 
Martin 1981).  Mrk~996 has 
a disk structure with a small scale length ($\sim$0.4 kpc) and spiral 
arms confined in the inner 160 pc. The faint diffuse envelope of this galaxy 
shows a distinct asymmetry, being more extended to the northeast side than to 
the southwest side. There is also an asymmetry in the spatial distribution of 
globular clusters around Mrk~996, where majority of them are found to the 
south of the galaxy. It has been considered as an indication of a tidal 
interaction or a past merger event (Thuan et al.\ 1996; hereafter Th96). The 
metallicity of Mrk~996 from its Oxygen abundance measurement has been estimated 
as $0.2$ solar ($12+\log$ (O/H)=8.0) by Th96. Mrk~996 has significantly larger 
N/O ratio than that in other BCDs having similar Oxygen abundances. This 
enhanced Nitrogen is explained by Pustilnik et al.\ (2004) as a result of some 
recent merger which caused the WR phase in the galaxy. 

The nebular \Ha emission from \HII regions is one of the fundamental 
indicators of the current($<5$~Myr) star-formation activity in galaxies 
(Kennicutt 1998; hereafter K98). Since the \Ha luminosity is proportional to 
the number of ionizing photons produced by hot and massive stars which is in 
turn proportional to their birth rate, the star formation rate (SFR) can be 
derived from the \Ha luminosity. As only young (lifetime $<20$~Myr), 
massive ($>10~M_\odot$) stars contribute significantly to the ionizing 
flux, the \Ha emission traces very recent star formation in galaxies. 
We note that the scaling between the \Ha flux and the SFR depends on the 
metallicity and the IMF of the stellar population. The calibration provided 
in K98 is for a Salpeter IMF and solar metallicity.
The nebular \Ha line being amongst the strongest emission 
lines in a galaxy, makes it as the preferred tracer for star formation in 
all types of galaxies at optical wavelengths. 
Narrow band \Ha photometric imaging is the best method to get an accurate 
estimate of the total \Ha flux. Slit spectroscopy on the other hand can not provide 
total \Ha flux as it normally samples a galaxy only over a small region. However, 
spectroscopic observations can provide vital information on line ratios which 
are essential to estimate extinction and temperature in the ionized regions. The 
\Ha flux in all cases needs to be corrected for Galactic and internal 
extinction, stellar continuum and other emission line contaminations in the 
passbands. These corrections are model based and require spectroscopic data.

\section{Observations and Results}

The observations were carried out in the \Ha narrow-band ($\lambda_c$ = 6570 
$\mathring{A}$,  FWHM $= 77 \mathring{A}$) and the SDSS r-band ($\lambda_c$ = 
6250 $\mathring{A}$, FWHM $= 1500 \mathring{A}$) using the 1.3 meter DFOT at 
Devasthal (longitude $= 79^{o}41'04''$~E, latitude $= 29^{o}21'40''$~N, 
altitude $\sim$2420~m above the mean sea level), near Nainital in the central 
Himalayan region in India. The main advantages of the Devasthal site are its 
dark sky, reasonably good $1''.2$ average seeing and low extinction (Sagar 
et al.\ 2010). A brief discussion on the telescope is presented in Sagar et al.\ 
2011. The telescope is equipped with low noise and fast readout Charge-
Coupled Devices (CCD) detectors and high transmission optical filters. This 
telescope uses the Ritchey-Chretien optical configuration with an f/4 Cassegrain focus 
providing a plate scale of $40''$ mm$^{-1}$. The telescope mount is of the 
fork-equatorial type, which requires rotation on only one axis while tracking 
celestial sources. The telescope can be pointed to a celestial object with 
an accuracy of $10''$  rms. The mechanical system provides a tracking accuracy 
of nearly $0''.5$ over 10-min for zenith angle 0 to 40 degree without an 
external guider. The data on Mrk~996 were recorded on a CCD camera having $2048
\times2048$ pixels, 13.5 micron pixel size, back-illuminated, and thermoelectrically 
cooled to -80\deg C. The CCD covers a field of view $18'\times18'$ on the sky. 
The CCD chip has QE of 90\% between $500-700$ nm and falls off to 50\% at 400 
and 900 nm. The CCD read noise is measured as 6.9~$e^{-}$ rms at 1 MHz read out 
rate with a gain of 2~$e^{-}/ADU$.

The observations were carried out on the dark (new moon) night of November 25, 
2011. These observations were made in the middle of servicing the telescope and 
therefore the optical system was not perfectly aligned. It might also be possible 
that many thermal sources inside the building were not minimized and hence image 
quality is not expected to be the best. Three \Ha frames of 600 sec each, five 
SDSS r-band frames of 300 sec each of the galaxy Mrk~996 and four frames (at 
different airmasses) of the standard spectrophometric star Feige~34 (Oke 1990) 
were recorded. The standard star is used to obtain the photometric calibration 
and also the atmospheric extinction parameters. 

CCD data reduction was performed using the package CCDRED in  the {\it 
Image Reduction and Analysis facility} (IRAF) software developed by 
{\it National Optical Astronomy Observatory}. The dark current in the CCD detector 
at $-80$\deg C is negligible ($ < 10^{-3}~e^{-}/sec/pixel$) compared to the sky 
background and hence dark frames were not recorded. After removing the cosmic 
ray events, all the aligned frames for each color filter were added to get the 
final image. The images were registered in the J2000 epoch equatorial coordinate 
system using the {\it Two Micron All Sky Survey} (2MASS) images. The 
underlying stellar continuum in the \Ha filter is subtracted using the 
standard procedure given by Waller (1990) and Spector et al.\ (2012). The Wide to 
Narrow Continuum Ratio (WNCR) is determined using non-saturated, non-variable field stars in the 
CCD image. The average WNCR is determined as $19.2 \pm 1.0$. The WNCR value is 
used to scale down the r-band image such that counts per second of field 
stars in the r-band and in the H$\alpha$-band are almost identical. Since both the \Ha and 
the r-band images were taken on the same night and in almost same seeing 
conditions, there was no need for any deconvolution. The continuum subtracted 
\Ha image is obtained by subtracting the scaled r-band image from the H$\alpha$
-band image. The aperture photometry of the star forming region in the continuum 
subtracted \Ha image was performed using the DAOPHOT package in IRAF. 

The underlying stellar continuum emission mapped by the r-band image and the 
ionized Hydrogen gas (star forming region) mapped by the continuum subtracted 
\Ha image are shown in Fig.~1. It can be seen that flux from majority of stars 
has been completely removed in the \Ha image. Residual flux can be seen around 
a few stars, which are confirmed as variable stars in the present analysis.
The FWHM on star images was found to be nearly 
$2''.4$. This FWHM is not a true representative of the atmospheric seeing as these 
observations were made in sub-optimal conditions in the observatory. The 
atmospheric extinction is determined as 0.15 mag airmass$^{-1}$ in the r-band. The sky 
brightness as determined from the SDSS r-band image is found to be $\sim$20.2 
mag arcsec$^{-2}$. 

\begin{figure}
\hspace{-0.9cm}
\includegraphics[height=9cm,width=9cm]{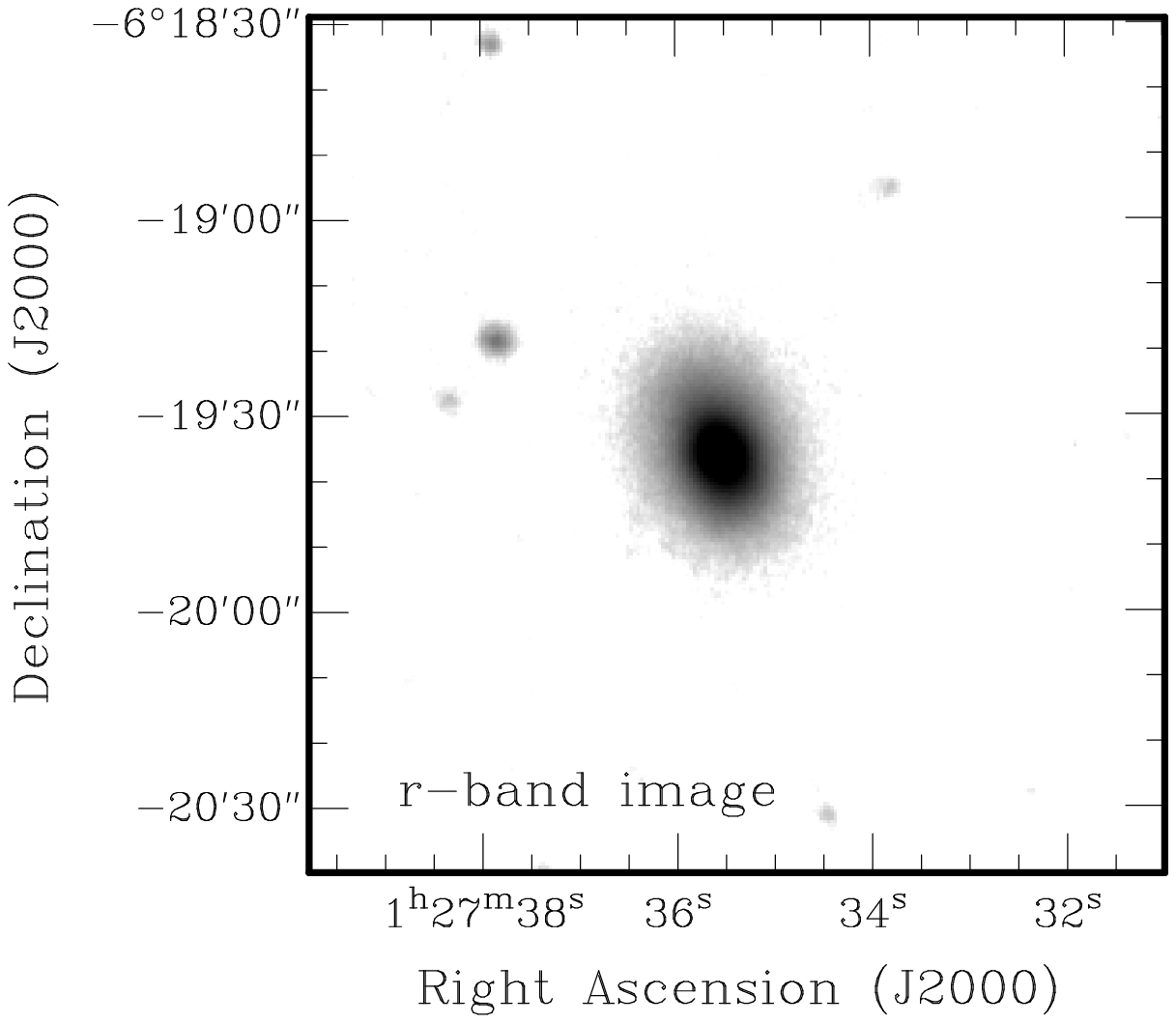}
    \begin{picture}(1,0)
        \put(-79,0){\includegraphics[height=9cm,width=9cm]{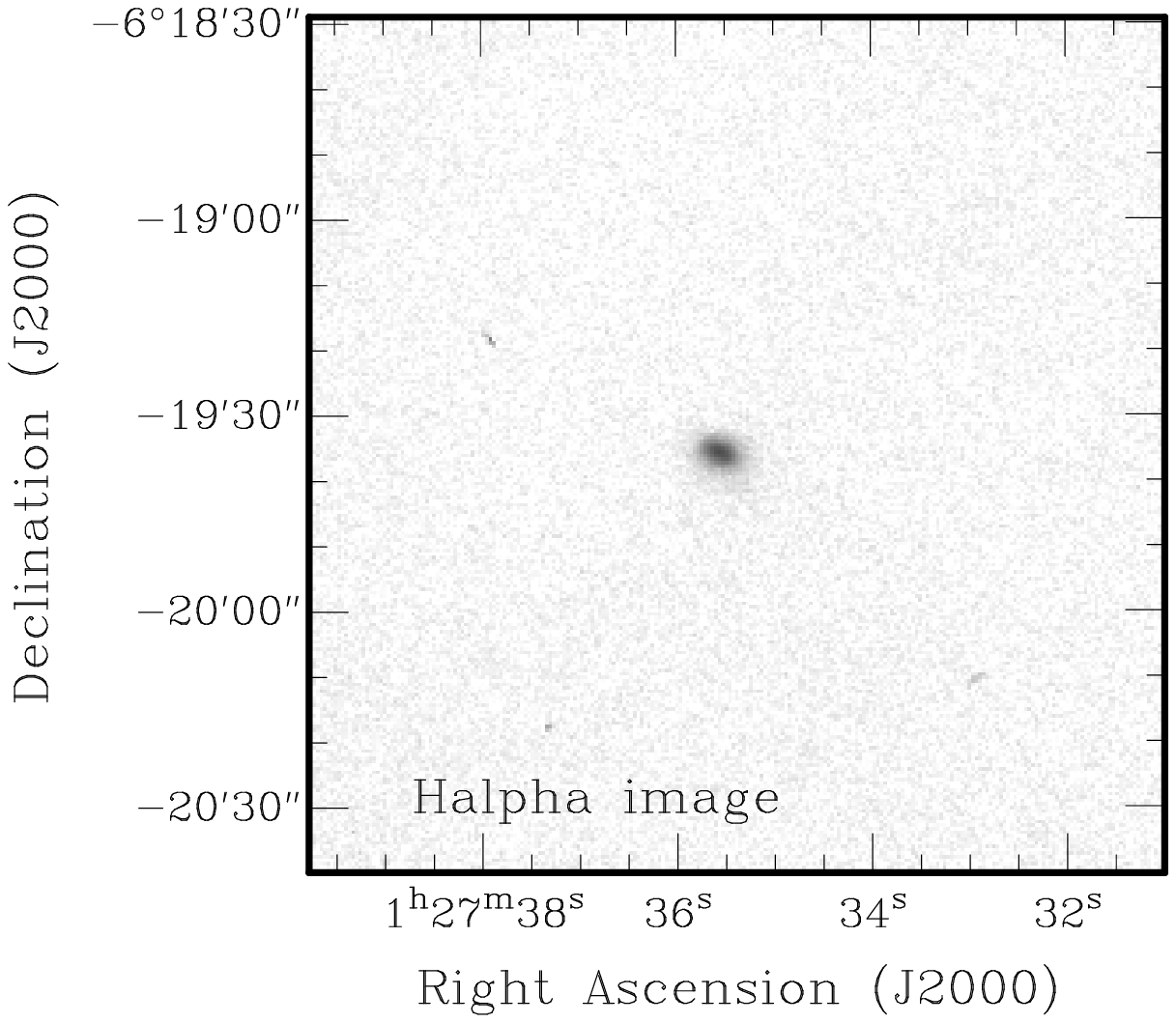}}
            \begin{picture}(1,0)
                \put(56,140){\includegraphics[width=1.7cm]{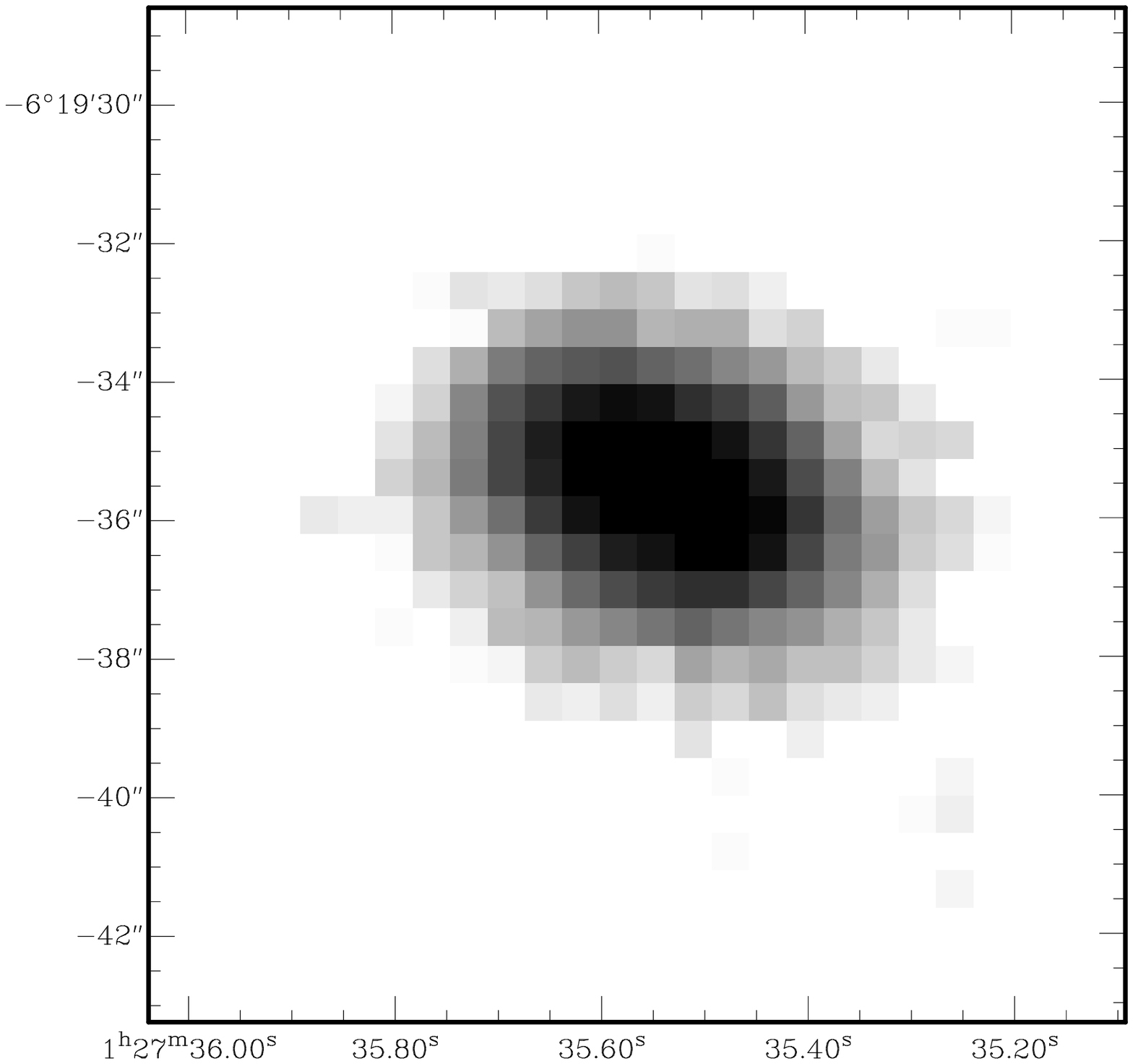}}
            \end{picture}
    \end{picture}
\vspace{-1.5cm}  
\caption{The SDSS r-band (left) and the continuum subtracted H$\alpha$ (right) 
image of Mrk~996 taken from the 1.3 meter telescope at Devasthal.}
\end{figure}

Using the calibrated spectrum of the standard spectrophotometric star, the 
mean instrumental response is estimated as $(3.3 \pm 0.3)\times 10^{-15}~
(erg~s^{-1}~cm^{-2})/(counts~s^{-1})$. 
This also includes the effect of the atmospheric extinction during the 
observations, and hence this value may vary slightly from night to night.
The \Ha surface brightness sensitivity ($3\sigma$) in 30 minutes of observations 
is obtained as $7.5\times10^{-17}~erg~s^{-1}~cm^{-2}~arcsec^{-2}$. This sensitivity 
is equivalent to an emission measure of $\sim$38$~pc~cm^{-6}$.

The calibrated \Ha flux is corrected for the narrowband filter 
transmission at the wavelength of \Ha emission from the galaxy. The \Ha 
line emission from Mrk~996 ($z=0.00541$) will be at 6598.5 $\mathring{A}$. 
Furthermore, for a telescope with a focal ratio number $n$, the central 
wavelength of a narrowband filter will be shifted towards the lower 
wavelength side by an amount $\Delta \lambda \approx \lambda_0/(16n^2\mu^2)$, 
where $\lambda_0$ is the central wavelength and $\mu$ is the effective 
refractive index of the interference filter. Therefore, for this telescope 
with a f-number 4 and taking the effective refractive index of the 
narrowband \Ha filter as 1.7, the shift in the central wavelength of the \Ha 
filter at 6570$\mathring{A}$ is found to be $\sim$9$\mathring{A}$.  The \Ha 
flux is therefore divided by $T_{N,H\!\alpha}= T_{N}(\lambda_{H\!\alpha,obs} + 
\frac{\lambda_0}{16n^2\mu^2})$ i.e. $\sim$0.56, where $T_{N,H\!\alpha}$ is 
the narrowband filter transmission at the \Ha emission wavelength from Mrk~996. 
This operation corrects the line flux from the decreased filter transmission 
at the \Ha wavelength corresponding to the redshift of the galaxy and the 
telescope focal ratio. The above correction is valid under the condition that 
the \Ha filter transmission curve is a Gaussian and the \Ha line emission from 
galaxy is narrow compared to the bandwidth of the filter. Both of these 
conditions are met in the present case. The raw \Ha flux from Mrk~996 is estimated 
as $(45 \pm 8)\times 10^{-14}~erg~s^{-1}~cm^{-2}$. This flux is in good 
agreement with that estimated by Gil de Paz et al. (2008), where \Ha flux was 
found as $(54 \pm 7)\times 10^{-14}~erg~s^{-1}~cm^{-2}$. Our estimate of the \Ha 
flux is further corrected for line contaminations, Milky-way extinction, 
internal extinction, stellar continuum and the underlying stellar absorption. 
These corrections are described in the next section.

The \Ha equivalent width, defined as the 
ratio of \Ha flux and specific continuum flux at the wavelength of 
\Ha emission, is obtained following the standard procedure 
given by Spector et al. (2012). The specific continuum flux is total continuum 
flux minus \Ha line 
contribution within the \Ha filter divided by its area ($1.07\times 
T_{N,max}\times FWHM_N$). The flux within the \Ha filter is derived from 
the aperture photometry of the \Ha band image using the same circular aperture 
as that used to get the \Ha line flux. Here, the underlying galaxy 
background in the narrowband image is estimated immediately next to that 
circular aperture. The \Ha equivalent width obtained using this method has 
been corrected for underlying stellar absorption by adding an expected 
value of \Ha stellar absorption equivalent width $EW^{abs}_{H_\alpha}$ 
following McCall et al. (1985). The \Ha equivalent width for Mrk~996 is 
estimated to be $(96 \pm 15) \mathring{A}$, in good agreement with the previous 
estimate of $\sim$109 $\mathring{A}$ by Gil de paz et al. (2008).

\section{Corrections to \Ha flux}

\subsection{Line contamination in filter passbands}

The narrowband (H$_\alpha$) filter contains significant flux from 
[N{\small II}] 6584 line. The wideband (r) filter, in addition, will contain 
significant flux from He{\small II} 6678, [S{\small II}] 6717 and 
[S{\small II}] 6731 emission lines apart from the \Ha 6563 line. It 
should be noted that the \Ha line itself is a contaminating line in the SDSS 
r-band for estimating continuum flux. The \Ha line flux  can be corrected 
for these line contaminations using the following relations over two steps:

\begin{equation}
F^i_{H\!\alpha} = \frac{F_{H\!\alpha}}{\left[1+\sum_{j}^{j\neq H\!\alpha} \left(\frac{T_N(\lambda_j)}{T_{N,H\!\alpha}} \right) \left(\frac{f_j}{f_{H\!\alpha}} \right)\right]}
\end{equation}

\begin{equation}
F^{ii}_{H\!\alpha} = F^i_{H\!\alpha} \left[1 + \frac{1}{WNCR} \sum_{j}^{}T_W(\lambda_j) \left(\frac{f_j}{f_{H\!\alpha}} \right) \right]
\end{equation}

\noindent where the flux ratios $f_j/f_{H\!\alpha}$ for each emission line 
$j$ in the wide-band filter are taken from the spectroscopic analysis of 
Mrk~996 by Th96. These ratios are provided in Table~1. Here, 
$T_N(\lambda_j)$ and $T_W(\lambda_j)$ are the transmissions in the 
narrowband (H$\alpha$) and the wideband (SDSS-r) filters respectively 
at the wavelength of emission line $j$. The eqs.~1 and 2 take care of 
line contaminations in \Ha and SDSS r-band filters respectively. The 
index superscript $i$ in flux values (F) is indicative of each step of 
correction applied to the \Ha flux. 

\begin{table}
\begin{center}
\caption{Flux ratio $f_j/f_{H\!\alpha}$ for various emission lines from Th96}
\begin{tabular}{cccc}
\hline
\hline
\bf {Line} & $\mathbf{f_j/f_{H\!\alpha}}$ & $T_N(\lambda_j)$ &  $T_W(\lambda_j)$\\
\hline

[\NII] 6584 & 0.0567 & 0.1 & 0.92\\

\HeI 6678 & 0.0142 & 0.001 & 0.92\\

[\SII] 6717 & 0.0197 & 0.0003 & 0.92\\

[\SII] 6731 & 0.0195 & 0.0003 & 0.92\\

\hline
\end{tabular} 
\end{center} 
\end{table}

The estimate of the \Ha flux is modified to $\sim$47$\times 10^{-14}~
erg~s^{-1}~cm^{-2}$, corrected for contaminating emission lines in the 
\Ha and SDSS-r passbands. Coincidentally, the modified value of flux is not too 
different from the un-corrected flux because (\Ha + other lines) contamination in 
the broadband filter is roughly similar to the line contamination in the \Ha filter. 

\subsection{Extinction correction}

The \Ha flux is decreased by extinction caused by the Milky-way.  
The extinction value is $A_r= 0.101$~mag in 
the direction of Mrk~996 in the SDSS r-band taken from the NASA/IPAC 
Extragalactic Database (NED) based on Schlafly \& Finkbeiner (2011) 
recalibration of the Schlegel, Finkbeiner \& Davis (1998) extinction 
map. The \Ha flux is modified to $\sim$51 $\times 10^{-14}~erg~s^{-1}
~cm^{-2}$ after this correction.

The \Ha flux also suffers from intrinsic extinction because of 
dust within the host galaxy. The \Ha flux can be corrected 
for the intrinsic extinction using the relation given by Lopez-Sanchez 
\& Esteban (2008) assuming Cardelli et al.\ (1989) 
extinction law with $R_V = 3.1$. In the present case, the internal 
extinction coefficient is $c(H_\beta) = 0.53$ 
using the Balmer line ratios taken from Th96. After applying 
this correction, the \Ha flux becomes $\sim$120 $\times 10^{-14}~
erg~s^{-1}~cm^{-2}$. It should be noted that the central region of 
Mrk~996 is highly obscured as evident from the $c(H_\beta)$ value.

\subsection{Continuum slope correction}

The continuum spectrum of Mrk~996 (c.f. Th96) is not flat within 
the bandwidths of SDSS-r and \Ha filters. The slope of 
the galaxy spectrum is such that the continuum emission 
within the r-band will be marginally greater than that in the \Ha band. 

It implies that in the continuum subtraction ($H\!\alpha - \frac{r}{WNCR}$) 
process, higher value of galaxy continuum has been removed from the \Ha band 
image, resulting in an underestimation of \Ha flux. If $s$ ($= \frac{1}{f_\lambda}\frac
{\Delta f_\lambda}{\Delta \lambda}$; $f_\lambda$ being the continuum flux at 
a particular wavelength) is the fractional decrease in continuum per unit 
wavelength increment, $\lambda_W$ is the central peak wavelength of 
wideband filter and $\lambda_{H\!\alpha}$ is red-shifted \Ha wavelength 
for the galaxy, then the corrected \Ha flux can be 
approximated by:

\begin{equation}
F^{v}_{H\!\alpha} = F^{iv}_{H\!\alpha} \left[1 + s(\lambda_{H\!\alpha}-\lambda_W)\right]
\end{equation}

The value of $s$ for Mrk~996 is found as $\sim$2 $\times 10^{-4}~
\mathring{A}^{-1}$. This correction brings the  \Ha flux to a value 
of $\sim$128 $\times 10^{-14}~erg~s^{-1}~cm^{-2}$.

\subsection{Underlying stellar absorption correction}

The massive stellar content (O, B, and A stars) in starburst galaxies dominate 
the continuum emission. These stars have strong Hydrogen Balmer absorption 
lines in their spectrum (Walborn \& Fitzpatrick 1990).  
The equivalent width of stellar Balmer absorption line is smaller than 
that of nebular emission line at the same wavelength, therefore  absorption 
features are completely masked in the optical spectra of galaxies 
(Gonzalez-Delgado et al.\ 1999). However, as the \Ha emission from ionized 
gas is superimposed on the stellar absorption line at the same wavelength, 
it makes the \Ha flux being underestimated. The \Ha flux can 
be corrected for the underlying stellar absorption using the following 
relationship (Hopkins et al.\ 2003): 

\begin{equation}
F^{vi}_{H\!\alpha} = F^{v}_{H\!\alpha} \left[1 + \frac{EW^{abs}_{H\!\alpha}}{EW^{obs}_{H\!\alpha}} \right]
\end{equation}

\noindent where the \Ha underlying stellar absorption equivalent width 
($EW^{abs}_{H\!\alpha}$) for H{\small II} galaxies is assumed as 
3$\mathring{A}$ (Gonzalez-Delgado et al.\ 1999) and the \Ha emission line 
equivalent width ($EW^{obs}_{H\!\alpha}$) is taken from this paper. This 
correction makes the final \Ha flux as $\sim$132 $\times 10^{-14}~
erg~s^{-1}~cm^{-2}$.

\subsection{Errors in estimates}

It is difficult to make an estimate of error in above corrections as 
uncertainties in flux corrections are largely unknown and model 
dependent. The instrumental (random) error makes the \Ha flux estimates 
to be accurate within nearly 18\%. Assuming that typical systematic error 
in each of the above corrections is of the order of 10\%, the error in 
the corrected estimates are nearly 22\%.  If both types of errors are 
added in quadrature, the final error can be taken as nearly 28\%.  We 
therefore quote the final corrected \Ha flux estimate for Mrk~996 from 
these observations and analyses as $(132 \pm 37)\times 10^{-14}~
erg~s^{-1}~cm^{-2}$. At a distance of ($D= 21.6\pm 0.1$~Mpc with 
$H_{0}=75$ km~s$^{-1}$~Mpc$^{-1}$) to the galaxy, the $H\!\alpha$ 
luminosity is $(7.4\pm2)\times10^{40}~erg~s^{-1}$.

\section{Discussions}

\subsection{Star Formation Rate in Mrk~996}

The SFR can be calculated from \Ha flux using the relation provided by K98:

\begin{equation}
\frac{SFR_{H\!\alpha}}{~M_\odot~yr^{-1}}=9.45\times 10^8 \left(\frac{D}{Mpc} \right)^2 \left(\frac{F_{H\!\alpha}}{erg/s/cm^2}\right)
\end{equation}

This relation holds good for a 
Salpeter IMF with mass limits of 0.1 and 100 $M_\odot$, and a stellar 
population with solar metallicity. Higher number of Lyman photons 
will be produced per unit stellar mass in low metallicity stars. 
Therefore, it is expected that K98 relation deviates in low metallicity 
systems such as dwarf galaxies, where SFR estimates based on K98 relation 
alone will be overestimated. Leitherer (2008) has estimated corrections to 
SFR in context of rotating and low metallicity massive stars. Based on 
these corrections, the star formation rate in Mrk~996 is predicted 
as $SFR_{H\!\alpha}= (0.4 \pm 0.1)~M_\odot$yr$^{-1}$. It should be noted 
that IMF has strong dependence on metallicity (cf. Zhang et al. 2007) 
in WR galaxies in the sense that higher number of massive stars are formed 
in low metallicity galaxies compared to that predicted from the Salpeter IMF. 
Therefore, overall SFR estimates using $H\alpha$ may require further corrections. 

The above estimates of SFR can be compared against other indicators of SFR 
such as the radio continuum luminosity at GHz frequencies. Mrk~996 is detected at 1.4 GHz in the 
Faint Images of Radio Sky at Twenty-centimeter (FIRST) survey images with a 
flux density of $0.5\pm0.3$ mJy. The radio emission is seen primarily from 
the center of galaxy. This radio flux is far lower than the non-thermal radio 
emission expected ($\sim$5 mJy) using Condon (1992) SFR-radio relationship 
for a normal star forming galaxy. We believe that the detected radio continuum in Mrk~996 is mainly 
thermal. This assumption is likely to be valid as supernova events have not 
taken place as evident from the presence of a large number of WR stars in the 
galaxy. Following the relation for Lyman continuum photons/SFR and thermal 
(free-free emission) radio flux (c.f. Caplan \& Deharveng 1986, Condon 1992, 
Hunt et al. 2004), the star formation rate is predicted to be nearly 
$0.3~M_\odot$~yr$^{-1}$, in good agreement with the overall SFR predicted 
from $H\alpha$ emission. It should be noted that compact and dense regions 
of star formation as in Mrk~996 may not be completely optically thin at 1.4 GHz 
and therefore SFR from radio continuum may be slightly underestimated.

\subsection{Radio-FIR correlation}

It is worthwhile to examine for Mrk~996 the radio-far infrared (FIR) 
correlation, known for normal star forming galaxies (Condon et al. 1991). 
Based on IRAS (Infra-Red Astronomical Satellite) detections at 60$\mu$m 
and 100$\mu$m with flux densities as 0.42 Jy and $<0.7$ Jy respectively, 
and the flux estimate at 1.4 GHz from the FIRST image (Sect 4.1), the 
ratio of FIR luminosity to radio luminosity, the $q$ parameter (cf. Yun 
et al. 2001) is $\sim$3 for Mrk~996. The average value of $q$ for normal 
galaxies is $<q>\sim2.3$. It implies that Mrk 996 deviates from the 
radio-FIR correlation and is significantly radio deficient. As discussed 
in the previous section, much of the radio continuum in Mrk~996 seems to 
be of thermal origin while that in normal star forming galaxies is mostly 
of the synchrotron origin, therefore the $q$ value once calculated for a 
synchrotron fraction alone would come out much higher for Mrk~996. Not many 
galaxies ($\sim$10) with $q\ge3$ (i.e. with the highest radio deficiencies) are known 
(Roussel et al. 2003, Omar \& Dwarakanath 2005). It is believed that high-q 
galaxies are seen at the very early phase of starburst following a long 
period of quiescence (Roussel et al. 2003). As Mrk~996 is a WR galaxy, the 
lack of supernova as discussed in the previous section is a plausible 
explanation for the deviation from the radio-FIR correlation.  However, as 
the central region in Mrk~996 is highly obscured and ionized, the free-free 
absorption at 1.4 GHz may be significant. In absence of any radio spectral 
measurement available to us, it is not possible to examine this aspect any 
further in this paper. A more detailed analysis of radio-FIR correlation 
using multi-frequency radio and infrared data at different bands (such as 
Spitzer data) will be very useful and interesting.

\subsection{Distribution of Ionized gas : evidence of minor merger}

The \Ha iso-intensity contours are overlaid on r-band contours 
in Fig.~2. It can be seen that \Ha emission is smoothly distributed 
without any significant clumps or knots. It is also evident from Fig.~2 
that the \Ha disk is mis-aligned with the continuum stellar disk.  
This misalignment has not been reported earlier. The 
misalignment of the two disks is nearly $40\hbox{$^\circ$}$. It is known that 
Mrk~996 has various asymmetries in its stellar continuum and globular 
cluster distribution, indicating a recent tidal interaction in the 
galaxy (Th96). It is quite possible that \Ha and stellar continuum 
disk misalignment is a result of tidal interaction where a fresh 
supply of gas from outer regions has fallen towards the center thereby 
causing a starburst. It has been seen in simulations (e.g., Hernquist 
\& Mihos 1995) that tidal interactions and minor mergers 
between galaxies can cause gas in the outer region to loose angular 
momentum and fall towards the centre of galaxy. This fresh supply of 
gas can be effectively converted to stars causing nuclear starburst 
(Mihos \& Hernquist 1994). As the new gas can come with different 
angular momentum, it is not necessary that orbits of fresh gas are 
aligned with that of the old stellar disk (Eliche-Moral et al.\ 2011, Haynes 
et al.\ 2000). Therefore, we believe that the current phase of star-burst 
in Mrk~996 has resulted from a recent tidal encounter, most likely a 
minor merger with a low-mass companion. 

\begin{figure}
\centering
\epsfig{figure= 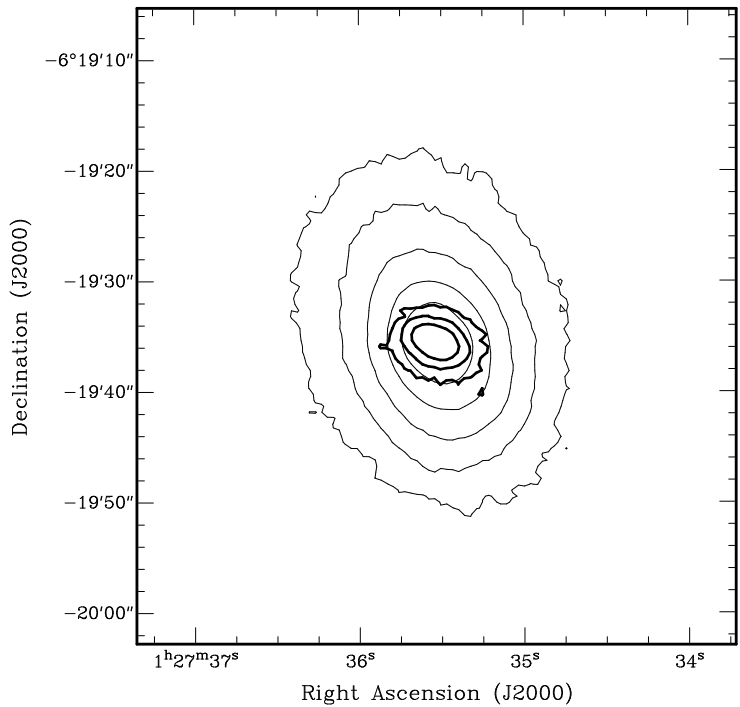,height=4in,width=4in,angle=0}
\caption{The H$_\alpha$ iso-intensity contours (thick line) overlying on 
SDSS r-band iso-intensity contours (thin line) showing the misalignment of 
the two disks, which may be because of tidal interaction disturbance. The 
contour levels are in logarithmic scale.}
\end{figure}

\section{Conclusions}

\Ha observations of a WR galaxy Mrk~996 have been reported here using the 
recently completed 
1.3~meter telescope at Devasthal. The instrumental response with \Ha 
filter  is estimated as $(3.3 \pm 0.3)\times 10^{-15}~(erg~s^{-1}~
cm^{-2})/(counts~s^{-1})$ at atmospheric extinction of 0.15 mag/airmass. 
The ratio of SDSS r-band to \Ha filter response is estimated as 
$19.2\pm 1.0$. The \Ha image presented here reach surface brightness 
sensitivity ($3 \sigma$) of $7.5\times10^{-17}~erg~s^{-1}~cm^{-2}~
arcsec^{-2}$, equivalent to an emission measure of $\sim$38~$pc~cm^{-6}$ 
at $T_e = 10000~K$ in nearly 30 minutes of observation. The main 
conclusions on the WR galaxy Mrk~996 from these observations are the 
following ---

\begin{itemize}

\item The \Ha flux from Mrk~996 is estimated as $(132 \pm 37)\times 
10^{-14}~erg~s^{-1}~cm^{-2}$, corrected for extinction, line contaminations 
in the passbands and underlying stellar absorption.

\item The SFR from \Ha flux corrected for the sub-solar metallicity of Mrk~996 
is estimated as $(0.4 \pm 0.1)~M_\odot$~yr$^{-1}$. This estimate is in good 
agreement with that indicated from the 1.4 GHz radio continuum emission, which 
is assumed to be entirely thermal.

\item Mrk~996 deviates from the radio-FIR correlation known for normal star forming galaxies. It is a radio deficient galaxy with a 'q' value of $\sim$3.

\item The \Ha emission is seen in a disk shape, misaligned by nearly $40\deg$ 
from its old stellar disk. It implies fresh supply of gas to center is most 
likely due to a recent tidal interaction event. This tidal interaction is 
likely to be responsible for the WR phase in the galaxy. 

\end{itemize}

\section*{Acknowledgments} 

SJ thanks Chrisphin Karthick for helps in the CCD data analysis. 
We thank Kuntal Misra for a careful reading of the manuscript and 
providing useful suggestions. We thank the referees for useful and 
constructive comments.
This research has made use of the NASA/IPAC Extragalactic Database (NED) 
which is operated by the Jet Propulsion Laboratory, California Institute 
of Technology, under contract with the National Aeronautics and Space 
Administration. This research has made use of NASA's Astrophysics Data 
System. We thank the staff of ARIES, whose dedicated efforts made these 
observations possible. DFOT is run by Aryabhatta Research Institute of 
Observational Sciences with support from the Department of Science and 
Technology, Govt. of India.

\end{document}